\PassOptionsToPackage{table}{xcolor}
\documentclass{article}
\usepackage{graphicx} 
\usepackage{amsmath}

\usepackage{iclr2025_conference}

\usepackage[utf8]{inputenc} 
\usepackage[T1]{fontenc}    
\usepackage{url}            
\usepackage{booktabs}       
\usepackage{amsfonts}       
\usepackage{nicefrac}       
\usepackage{microtype}      
\usepackage{float}
\usepackage{caption}

\usepackage{booktabs}
\usepackage{multirow}
\usepackage{siunitx}
\usepackage{changepage}
\usepackage{longtable}
\usepackage{textgreek}
\usepackage{subcaption} 

\iclrfinalcopy
\newtoggle{showtodo}
\toggletrue{showtodo}

\title{Exploring zero-shot structure-based protein fitness prediction}

\author{Arnav Sharma$^{* \dagger}$ \& Anthony Gitter$^{* \dagger \ddagger}$\\
$^{*}$ Department of Computer Sciences, \\
$^{\ddagger}$ Department of Biostatistics and Medical Informatics \\
University of Wisconsin-Madison \\
$^{\dagger}$ Morgridge Institute for Research \\
\texttt{arnav@cs.wisc.edu, gitter@biostat.wisc.edu}
}

\begin{document}

\maketitle

\begin{abstract}
The ability to make zero-shot predictions about the fitness consequences of protein sequence changes with pre-trained machine learning models enables many practical applications.
Such models can be applied for downstream tasks like genetic variant interpretation and protein engineering without additional labeled data. The advent of capable protein structure prediction tools has led to the availability of orders of magnitude more precomputed predicted structures, giving rise to powerful structure-based fitness prediction models. Through our experiments, we assess several modeling choices for structure-based models and their effects on downstream fitness prediction. Zero-shot fitness prediction models can struggle to assess the fitness landscape within disordered regions of proteins, those that lack a fixed 3D structure.
We confirm the importance of matching protein structures to fitness assays and find that predicted structures for disordered regions can be misleading and affect predictive performance.
Lastly, we evaluate an additional structure-based model on the ProteinGym substitution benchmark and show that simple multi-modal ensembles are strong baselines.
\end{abstract}

\section*{Introduction}
Designing proteins with desired properties carries immense promise with applications in therapeutic development, agriculture, chemical manufacturing, and biofuels \citep{liu_chapter_2022}. Effectively navigating the large design space of proteins requires understanding the relation between a protein's sequence and its function \citep{romero_navigating_2013}. The opportunities in this scientific domain and excitement around protein science complemented by the rise in popularity of machine learning methods have driven a wide variety of new methods for  quantitative protein function prediction \citep{yang_machine-learning-guided_2019}.
Computational methods built using aligned sequences or multiple sequence alignments (MSAs) \citep{rao_msa_2021} for evolutionarily related proteins can estimate protein fitness scores based on conservation or the likelihood a new sequence belongs to the protein family.
In contrast, protein language models (PLMs) \citep{rives_biological_2021} train on and represent all known natural protein sequences to learn more general statistical patterns of evolution.

Protein structures provide additional information for estimating the fitness consequences of sequence changes.
The increase in quality of predicted protein structures from models like AlphaFold 2 \citep{jumper_highly_2021} has been followed by an increasing number of structure-based models for predicting protein properties \citep{dauparas_robust_2022, hsu_learning_2022, blaabjerg_rapid_2023}.
The initial ProteinGym 0.1 \citep{notin_tranception_2022}, a benchmark for quantitative protein fitness prediction that used deep mutational scanning (DMS) assays for evaluation, did not contain structure-based models. However, ProteinGym 1.0 \citep{notin_proteingym_2023} was released with three inverse folding methods that use structures: ESM-IF1 \citep{hsu_learning_2022}, ProteinMPNN \citep{dauparas_robust_2022}, and MIF \citep{yang_masked_2023}. 
Additional structure-based models were added after the ProteinGym publication.
This warrants a closer examination of structure-based models' behavior and protein fitness prediction performance to determine what modeling choices and biological factors affect their predictions.

Here we explore some design choices around using structure-based models for protein fitness prediction using the ProteinGym benchmark \citep{notin_proteingym_2023}. We examine how the type of protein structure (predicted or experimental) can affect model predictions. We observe that a number of assays in the ProteinGym benchmark involve proteins containing intrinsically disordered regions (IDRs). These are regions in the protein sequence that lack any rigid 3D structure and can adopt multiple varied conformations \citep{ijms232214050}. We demonstrate how IDRs can affect model predictions made by various models.
Finally, we expand the structure-based models in ProteinGym, comparing them to strong existing MSA- and PLM-based models.
We show strengths of uni-modal and multi-modal structure-based models, especially for stability assays.
Additionally, we show that simple ensemble models that include structure perform surprisingly well compared to more sophisticated models.

\section*{Results}
We use ProteinGym \citep{notin_proteingym_2023} to explore structure-based predictions of protein fitness.
ProteinGym hosts a collection of DMS substitution assays measuring quantitative function of proteins belonging to different taxonomies and five different function types: activity, binding, expression, organismal fitness, and stability (Table \ref{tab:count_table}). Furthermore, the benchmark hosts a variety of models sorted according to their average Spearman correlation with measured function across protein and function types.

\subsection*{Choice of structure affects zero-function protein function prediction}
Previous work on PLMs has shown that given a corrupted protein sequence, predicting log-likelihood of mutations at the corrupted residue can be indicative of protein fitness \citep{rives_biological_2021, yang_convolutions_2024}. Unlike sequence-based PLMs, inverse folding models take a corrupted sequence as well as the backbone structure of a protein and try to predict the likelihood of the corrupted residue. This explicit conditioning on protein structure has shown to improve recovery of the corrupted residues \citep{yang_masked_2023} and is therefore indicative of better fitness prediction. 

We first benchmark zero-shot predictive performance of the structure-based ESM-IF1 on all DMS substitution assays from ProteinGym for which we could obtain matching experimental structures. 
One key difference between experimental and predicted structures might be that predicted structures only contain the coordinates of a target chain whereas an experimental structure might contain a protein complex with multiple chains. Therefore, in order to explicitly account for this difference, we separate the results according to whether the experimental structure is a monomer (has a single chain) or a multimer (has multiple chains).
We compare the difference of Spearman correlation using predicted versus experimental structures.
Previous work has shown that experimental structures improve prediction of protein properties such as Enzyme Commission number (EC) and Gene Ontology term (GO) relative to structures predicted by AlphaFold 2 \citep{huang_protein_2024}.
Our results show that, for most assays in ProteinGym with experimental structures, Spearman correlation achieved using AlphaFold 2 predicted structures tends to be higher than that achieved using experimental structures (\tablename{ \ref{tab:delta_rho_table}}). A more detailed breakdown can be found in \figurename{~\ref{fig:spearman_diff}} in the Appendix.

\begin{table}[htb]
    \begin{tabular}{lccc} 
            \toprule
            \multirow{1}{*}{\textbf{Structure Type}} &
            \textbf{$\rho_{\text{pred}} - \rho_{\text{exp}} \ge 0$} & \textbf{$\rho_{\text{pred}} - \rho_{\text{exp}} < 0$}  & 
            \textbf{$\%$ with $\rho_{\text{pred}} - \rho_{\text{exp}} \ge 0$}\\
            \midrule
            \textbf{Monomers} &
            {38} & {13} & {74.5$\%$}\\
            \textbf{Multimers} & {8} & {2} & {80$\%$} \\
            \bottomrule
    \end{tabular}
    \caption{Count of DMS assays in ProteinGym with positive and negative $\rho_{\text{pred}} - \rho_{\text{exp}}$}
    \label{tab:delta_rho_table}
\end{table}

\subsection*{Intrinsically disordered regions are prevalent in ProteinGym and inform structure-based model performance}

52 of the 186 (28\%) unique UniProt IDs in ProteinGym are proteins that are annotated as having disordered regions in DisProt in regions of the sequence covered by the DMS assay (\tablename{ \ref{tab:disorder_content}}).
These proteins are associated with 63 of the 217 (29\%) ProteinGym DMS substitution assays. Their disorder content ranges from 1.45\% to 100\% with a median of 15.67\% and a mean of 23.56\%. Nine proteins have over 45\% disorder content. The Appendix
details our process of identifying disordered regions in ProteinGym assays.

We assess the effect disordered regions have on function prediction at the protein and residue-level.
We first benchmark models in \tablename{ \ref{tab:spearman_table}} as well as ESM2 650M \citep{lin_evolutionary-scale_2023} on 63 DMS assays that have disordered regions in the target sequence defined by ProteinGym (\tablename{ \ref{tab:disorder_content}}).
There is some variation in performance across models, which may be related to the models' different input modalities (\figurename{ \ref{fig:prot_level_order_vs_dirsorder_spearman}}). 

In order to more directly assess how disordered regions affect zero-shot fitness prediction quality, we examine the subset of 43 DMS assays that contain mutations in both disordered and ordered regions corresponding to 36 proteins (\tablename{s \ref{tab:disorder_content} and \ref{tab:count_disordered_proteins}}).
Disordered regions affect not only structure-based models like ESM-IF1 (\figurename{ \ref{fig:order_vs_dirsorder_spearman:esm_if}}), but also multi-modal models like ProtSSN \citep{tan_semantical_2024} (\figurename{ \ref{fig:order_vs_dirsorder_spearman:protSSN}}), SaProt \citep{su_saprot_2023} (\figurename{ \ref{fig:order_vs_dirsorder_spearman:saProt}}), and TranceptEVE L \citep{notin_trancepteve_2022} (\figurename{ \ref{fig:order_vs_dirsorder_spearman:tranceptEVE}}) for most function types.  Disorder in proteins can also be detrimental to predictions made by PLMs (\figurename{ \ref{fig:order_vs_dirsorder_spearman:esm2}, \ref{fig:order_vs_dirsorder_spearman:vespa}, \ref{fig:esm2_order_vs_dirsorder_p53}}). We also observe this phenomenon in ensembles of different models (\figurename{ \ref{fig:order_vs_dirsorder_spearman:struct2seq}, \ref{fig:order_vs_dirsorder_spearman:ensemble2}, \ref{fig:order_vs_dirsorder_spearman:ensemble3}, \ref{fig:order_vs_dirsorder_spearman:ensemble4}}; ensembles defined in \tablename{ \ref{tab:spearman_table}).
Disordered regions tend to be fast-evolving and less conserved \citep{brown_evolutionary_2002, light_protein_2013, fawzy_assessing_2025}, which may partially explain this observation.

Because many structure-aware protein function prediction methods rely on structures predicted by models like AlphaFold 2, studying the consequences of disordered regions on predicted structures can inform fitness prediction outputs.
We take a more in-depth look at two such proteins with disordered regions: P37840 (human \textalpha-synuclein) and Q99801 (human NKX3-1). \textalpha-synuclein is a protein associated with Parkinson’s disease that can take on disordered, helical, or \textbeta-sheet-rich aggregate conformations \citep{newberry_deep_2020, newberry_robust_2020}.
DMS data \citep{newberry_deep_2020, newberry_robust_2020} support that an \textalpha-helicial membrane-bound conformation with increasing disorder toward the C-terminal end is relevant for \textalpha-synuclein toxicity.
Comparing experimental and predicted \textalpha-synuclein structures (\figurename{ \ref{fig:1XQ8_alignment}}), neither is a perfect match for this membrane-bound conformation, but the helical experimental structure is much more similar.
The substantial difference between the two structures likely explains why this DMS assay SYUA\_HUMAN\_Newberry\_2020 has the most negative $\rho_{\text{pred}} - \rho_{\text{exp}}$ in \figurename{ \ref{fig:spearman_diff}}.

NKX3-1 has a small positive $\rho_{\text{pred}} - \rho_{\text{exp}}$ for the NKX31\_HUMAN\_Tsuboyama\_2023\_2L9R DMS assay (\figurename{~\ref{fig:spearman_diff}}).
Although the protein contains disordered regions, closer examination shows that the DMS assay only spans 61 positions of the protein that are ordered.
The experimental and predicted structures cover this same region of the protein and are quite similar (\figurename{~\ref{fig:2L9R_alignment}}), consistent with the minimal difference in DMS prediction performance.

\subsection*{Multi-modal models are better zero-shot predictors}

Besides the choice of input structure, another aspect that affects the performance of structure-based models is how and whether the structural information is combined with other data modalities.
Both MSAs \citep{marks_protein_2011} and PLM embeddings \citep{rives_biological_2021, zhang_protein_2024} already provide some information about protein structure.
We assess whether multi-modal models that explicitly use protein structure input have different performance characteristics on the protein fitness prediction task than models that implicitly contain structural information via MSA, PLM, or both inputs.
We create ensembles of strong existing zero-shot models by choosing the best-performing sequence- and MSA-based models from ProteinGym and combining each with the ESM-IF1 structure-based model to assess how adding the structure input affects predictive performance.

Benchmarking strong models for different input modalities as well as ensembles using these models on the ProteinGym benchmark shows that combining different modalities (through ensembling or using joint architectures) exhibits some of the highest metrics (\tablename{s \ref{tab:spearman_table}} and \ref{tab:recall_table}).
Models that incorporate protein structure information tend to perform well on DMS substitution assays measuring stability, consistent with previous results \citep{paul_combining_nodate, blaabjerg_joint_2023}. 
Furthermore, we see that even simple ensembles of models combining different modalities achieve some of the highest values across both Spearman correlation and Top 10 recall. This suggests that multi-modal models seem to capture complementary predictive signals that models operating on a single modality do not. Furthermore, multi-modal models such as ProtSSN, SSEmb, TranceptEVE L, and SaProt may still underperform simpler ensembled methods like StructSeq \citep{paul_combining_nodate} and others.

\begin{table}[htbp]
    \begin{adjustwidth}{-2cm}{}
        \begin{tabular}{@{\hspace{-0.3cm}} l c cccccc @{\hspace{-0.7cm}} c} 
            \toprule
            \multirow{2}{*}{\textbf{Model}} &
            \multirow{2}{*}{\textbf{Modality}} &
            \multicolumn{5}{c}{\textbf{Avg. $\rho \;$ by Assay Function}} &
            \multirow{2}{*}{\textbf{Avg. $\rho$}} \\
            & &  &  &  &  & \\
            & & {Activity} & {Binding} & {Expression} & {Organismal} & {Stability} \\
            & & {} & {} & {} & {Fitness} & {} \\
            \midrule
            ESM-IF1$^{*}$ & Structure & 0.368 & \textcolor{blue!90}{0.389} & 0.407 & 0.331 & \textcolor{olive!90}{0.624} & 0.424 & \\
            GEMME$^{*}$ & MSA & 0.482 & 0.383 & 0.438 & \textcolor{blue!90}{0.453} & 0.519 & 0.455 & \\
            VESPA$^{*}$ & PLM & 0.468 & 0.366 & 0.404 & 0.442 & 0.500 & 0.436 & \\
            ProtSSN (ensemble) $^{*}$ & Structure + PLM & 0.466 & 0.366 & 0.449 & 0.397 & 0.568 & 0.449 & \\
            SSemb & Structure + MSA & 0.436 & 0.386 & 0.469 & 0.422 & 0.596 & \textcolor{blue!90}{0.462} & \\
            TranceptEVE L$^{*}$ & MSA + PLM & \textcolor{olive!90}{0.487} & 0.376 & 0.457 & \textcolor{red!90}{0.460} & 0.500 & 0.456 & \\
            SaProt (650M)$^{*}$ & Structure + PLM & 0.458 & 0.379 & \textcolor{olive!90}{0.488} & 0.373 & 0.592 & 0.458 & \\
            \midrule
            Ensemble 1  & Structure + & & &  &  & & & \\
            StructSeq & MSA + PLM & \textcolor{blue!90}{0.485} & \textcolor{red!90}{0.432} & \textcolor{blue!90}{0.483} & 0.439 & \textcolor{red!90}{0.639} & \textcolor{red!90}{0.495} & \\ 
            \midrule
            Ensemble 2  & Structure + & & &  &  & & & \\
            ESM-IF1, Tranception & MSA + PLM & {0.479} & \textcolor{olive!90}{0.418} & \textcolor{red!90}{0.491} & {0.438} & \textcolor{blue!90}{0.621} & \textcolor{olive!90}{0.489} & \\
            \midrule
            Ensemble 3  & Structure + & & &  &  & & & \\
            ESM-IF1, GEMME & MSA & \textcolor{red!90}{0.491} & {0.387} & {0.440} & \textcolor{olive!90}{0.454} & {0.529} & {0.460} & \\
            \midrule
            Ensemble 4  & Structure + & & &  &  & & & \\
            ESM-IF1, VESPA & PLM & \textcolor{blue!90}{0.485} & {0.372} & 0.416 & {0.449} & 0.558 & {0.456} & \\
            \bottomrule
        \end{tabular}
        \caption{Spearman correlation coefficient ($\rho$) of model predictions versus DMS assay scores according to the function type of each DMS assay. Scores for model names annotated with an asterisk ($^{*}$) were calculated using their predictions provided in ProteinGym \citep{notin_proteingym_2023}. Scores highlighted in red, olive, blue represent the \textcolor{red!90}{1$^\text{st}$}, \textcolor{olive!90}{2$^\text{nd}$}, and \textcolor{blue!90}{3$^\text{rd}$} highest score for each group. See \figurename{ \ref{fig:spearman_distribution} for score distributions.}}
        \label{tab:spearman_table}
    \end{adjustwidth}
\end{table}

\section*{Methods}
We obtain DMS data and AlphaFold 2 predicted protein structures from ProteinGym \citep{notin_proteingym_2023}, experimental protein structures from the RCSB Protein Data Bank (PDB) \citep{zardecki_rcsb_2016}, and disordered protein annotations from DisProt \citep{aspromonte_disprot_2024}. Predicted monomer structures in the ProteinGym were generated using AlphaFold 2 v2.3.1 with full\_dbs and default parameters \citep{notin_proteingym_2025}.
Additional dataset details are in the Appendix.

There are many protein fitness prediction algorithms included in ProteinGym and far more not integrated into the benchmarking framework.
We prioritized representative alignment-based, PLM-based, and structure-based models that performed well on the ProteinGym zero-shot DMS substitution assays as well as additional structure-based models and ensembles.
The specific models and ensembles are described in the Appendix.

In order to evaluate the performance of each model on the ProteinGym dataset, we consider two metrics:
Spearman correlation ($\rho$) and Top 10 Recall (Appendix). The average score for each metric follows the convention established by ProteinGym. We first take the average of scores belonging to assays targeting the same protein (determined by UniProt IDs). We then take the average of the per protein averages across assay functions. Finally, the average column in \tablename{s \ref{tab:spearman_table}} and \ref{tab:recall_table} represents a mean of per function average scores. 

\section*{Discussion}
Our goal is not to assert that one algorithm is the best overall for zero-shot protein fitness prediction.
Rather, we provide insights into the strengths and weaknesses of different algorithms, input protein representations, and their combinations. Although MSA-based models \citep{marks_protein_2011} and PLMs \citep{rives_biological_2021, zhang_protein_2024} implicitly contain information about protein structure, explicit structure-based models do in fact benefit in zero-shot prediction performance on the ProteinGym DMS substitution assays. This finding is further bolstered by observing how combining different modalities can lead to improved predictive performance. In addition, structure-based models using predicted structures as inputs often show better performance on ProteinGym's zero-shot prediction tasks.
We note that ESM-IF is explicitly trained on AlphaFold 2 predicted structures \citep{hsu_learning_2022}, which could explain why predicted structures outperform experimental ones in our experiments.

Another reason for the discrepancy between model performance on predicted and experimental structures could be the lack of manual validation of our chosen structures. As detailed in the Appendix, we may have selected experimental structures misaligned with the fitness prediction assays we benchmark in \figurename{ \ref{fig:spearman_diff}}. Due to the presence of proteins bound to other molecules, alternative conformations, and protein modifications made to facilitate structure characterization in the PDB, experimental structures can be misaligned with a functional property of interest. However, this problem---as evident in \figurename{ \ref{fig:1XQ8_alignment}}---can also be observed in predicted structures. Our finding highlights the need for matching structures with downstream properties when making zero-shot fitness predictions and interpreting structure prediction models like AlphaFold 2 to understand how a protein's structure was predicted. Lastly, \citet{rives_biological_2021, yang_convolutions_2024} found that masked language modeling is highly indicative of protein fitness, but this may not be the case for GO and EC prediction tasks used by \citep{huang_protein_2024}.

The best function prediction algorithm can depend on the protein and function (Tables~\ref{tab:spearman_table} and~\ref{tab:recall_table}).
For some functions such as non-native enzymatic activity, zero-shot predictions are uncorrelated with functional assays \citep{yang_active_2024}.
Furthermore, the ProteinGym assays represent only a fraction of available DMS assays, and our results may change across a larger set of assays. 
ProtaBank \citep{wang_protabank_2018} and MaveDB \citep{rubin_mavedb_2022} catalog additional DMS datasets, and more are published continually.

One limitation of our study is that we did not use the same MSAs for all models. To get the best representation for SSEmb \citep{blaabjerg_joint_2023} on zero-shot prediction tasks, we generated MSAs using the process chosen by its authors instead of using ProteinGym MSAs. Furthermore, despite the zero-shot setting being a useful paradigm, it is also important to study how these models perform under supervised fine-tuning. Considering only MSAs, structure, and sequence as input modalities leaves out models like METL \citep{gelman_biophysics-based_2024} and RaSP \citep{blaabjerg_rapid_2023} that are trained on alternate modalities like protein stability and biophysical simulations. We could only analyze a small subset of DMS assays with their experimental structures in part due to missing residues in the experimental structures.
We could expand this subset by relaxing our criteria for selecting experimental structures to allow a small fraction of missing residues and model the missing residues with Rosetta \citep{huang_rosettaremodel_2011, song_high-resolution_2013, leman_macromolecular_2020}. 
In addition, we only considered a single experimental structure per protein and predicted structures from a single algorithm.
Further exploring the stability DMS assays, which are mostly cDNA display proteolysis assays \citep{tsuboyama_mega-scale_2023}, may explain why the two structure-based models do not see same drop in performance for stability assays that they do in other function types.

Our preliminary examination of intrinsically disordered regions in ProteinGym proteins shows how disorder can affect fitness predictions of not just structure based models, but those based on sequences and MSAs. For example, \figurename{ \ref{fig:esm2_order_vs_dirsorder_p53}} shows ESM2 650M predictions for the P53\_HUMAN\_Giacomelli\_2018\_Null\_Etoposide assay separated by mutations in ordered and disordered regions.
In the disordered region, ESM2 predicted pseudo log-likelihood correlates less with DMS scores, and the masked marginal probabilities have a slightly smaller range.
Our results on DMS data are consistent with AlphaMissense's lower performance for pathogenic variant prediction within disordered regions \citep{cheng_accurate_2023} and reduced sensitivity in a systematic evaluation of pathogenicity prediction \citep{fawzy_assessing_2025}.
Out of the structure-based models that we consider, ProtSSN \citep{tan_semantical_2024} and SSEmb \citep{blaabjerg_joint_2023}, do not give special treatment to disordered regions of a protein or low confidence regions of a predicted structure. ESM-IF1 \citep{hsu_learning_2022} on the other hand, takes the confidence of AlphaFold 2 \citep {jumper_highly_2021} predicted structures into account by masking low pLDDT coordinates. SaProt \citep{su_saprot_2023} also takes confidence of predicted structures into account by masking structure tokens corresponding to low pLDDT residues and only using sequence tokens.
Structure-based models are broadly applied for a variety of tasks beyond fitness prediction \citep{fout_protein_2017, gligorijevic_structure-based_2021, xia_graphbind_2021, gao_hierarchical_2023}, so our initial exploration of intrinsically disordered regions in ProteinGym presents a general opportunity for future model development. 
Future structure-based models could consider ways to account for the unique aspects of intrinsically disordered proteins or support conformational ensembles \citep{lindorff-larsen_potential_2021}.
Lastly, we see that even though multi-modal models capture a more holistic representation of proteins, simple ensembles of different models can often surpass their performance. This motivates further research into architectures and training objectives that better combine information across modalities as well as using simple multi-modal ensembles as baselines when developing those models.

\subsection*{Acknowledgments}
We thank Sam Gelman and Bryce Johnson for feedback and modeling suggestions as well as Moses Milchberg for protein structure discussions.
This research was supported by National Science Foundation awards CHE 2226451 and OAC 2030508 and computing resources at the University of Wisconsin-Madison Center for High Throughput Computing \citep{chtc}.

\subsection*{Code availability}
Code to reproduce our results can be found in our GitHub repository
(\url{https://github.com/gitter-lab/benchmarking-structure-based-models}) 
and is archived on Zenodo at 
(\url{https://zenodo.org/doi/10.5281/zenodo.13821572}). 
MSAs, structures, and additional results are available in the Zenodo repository 
(\url{https://doi.org/10.5281/zenodo.13819823}).

Most of our datasets come from the ProteinGym benchmarks and can be downloaded from \url{https://proteingym.org/download}. As indicated, we used the zero-shot model scores provided by ProteinGym to compute metrics for models that are already part of the benchmark. See the Appendix for details.

\clearpage
\bibliographystyle{iclr2025_conference}
\bibliography{iclr2025_conference}

\clearpage

\appendix

\section*{Appendix}
\subsection*{Datasets} \label{app:subsec:Datasets}
ProteinGym \citep{notin_proteingym_2023} is one of the most comprehensive datasets available for evaluating zero-shot protein fitness prediction. It presents a collection of DMS substitution assays measuring various protein functions: activity, binding, expression, organismal fitness, and stability \citep{notin_proteingym_2023}. Furthermore, the dataset also contains different input data modalities for all the proteins.
We used 216 DMS substitution assays from ProteinGym (\tablename{ \ref{tab:count_table}}). We did not use BRCA2\_HUMAN\_Erwood\_2022\_HEK293T. This is because its structure is split across multiple files, and MSA generation for SSEmb proved to be time-consuming.

In order to compute scores for SSemb \citep{blaabjerg_joint_2023}, however, we had to use MMseqs2 \citep{steinegger_mmseqs2_2017, mirdita_colabfold_2022} to create new MSAs. We used the notebook provided by SSemb to serially generate an alignment for each protein in the ProteinGym DMS substitutions dataset by processing each pdb file from ProteinGym.

To get experimental structures for each assay in ProteinGym we queried the RCSB PDB \citep{zardecki_rcsb_2016}. Details on the query parameters and selection process can be found below. We ended up with 65 different assays from the ProteinGym benchmark with matching experimental structures. These assays are shown in \figurename{ \ref{fig:spearman_diff}}, and their functions are summarized in \tablename{ \ref{tab:count_table}}.
We aligned experimental and predicted structures with the RCSB PDB pairwise structure alignment tool \citep{bittrich_rcsb_2024} and the TM-align alignment \citep{zhang_tm-align_2005}.

We obtained the disorder content from the 2024\_12 DisProt release \citep{aspromonte_disprot_2024}.
We mapped the ProteinGym UniProt \citep{the_uniprot_consortium_uniprot_2023} entry names to IDs (that is, accession numbers), replacing the outdated PSAE\_SYNP2 with PSAE\_PICP2.
ANCSZ represents an ancestral sequence, not a UniProt ID, and could not be mapped, so we obtained 186 unique UniProt IDs.
We intersected these with all UniProt IDs from the DisProt database.

\subsection*{Protein fitness prediction algorithms}

Score for algorithms marked with an asterisk ($^*$) were taken from the ProteinGym benchmark. These scores can be obtained by downloading the zero shot substitution scores provided on \url{https://proteingym.org}. For all other models, scores were generated manually by running inference on the ProteinGym zero shot substitution benchmark.

\begin{itemize}
    \item \textbf{VESPA$^*$ \citep{marquet_embeddings_2022}:} This model uses protein sequence as an input. It forgoes the need for MSA by using the embeddings from a pretrained PLM along to compute conservation information along with log-probability odds for variants to estimate their fitness effect.
    \item \textbf{GEMME$^*$ \citep{laine_gemme_2019}:} This model operates on the multiple sequence alignment (MSA) of a protein as input. The key idea is to present sequences homologous to the query as a tree (by sorting on sequence similarities) and extracting conserved residues. This allows the model to predict protein functions based on evolutionary signal for each protein.  
    \item \textbf{ESM Inverse Folding$^*$ \citep{hsu_learning_2022}:} This model proposes the idea of using a structural encoder in addition to a sequence-to-sequence transformer model in order to consume a protein's folded structure and corrupted sequence to predict the original sequence from it. We chose this model instead of ProteinMPNN \citep{dauparas_robust_2022} because both the ProteinGym benchmark \citep{notin_proteingym_2023} as well as prior work \citep{paul_combining_nodate} have noted that ESM inverse folding provides a stronger baseline representation for structure-based models.
    \item \textbf{ProtSSN$^*$ \citep{tan_semantical_2024}} This model uses both protein sequence and structure as input. The model uses a graph neural network to process a protein's structure graph representation. The embeddings from a PLM are used as initial node embeddings and processed over the structure graph.
    \item \textbf{TranceptEVE L$^*$ \citep{notin_trancepteve_2022}:} TranceptEVE combines signal predictive of protein function from the protein's MSA (using EVE and Tranception's MSA retrieval \citep{notin_tranception_2022}) well as sequence (Tranception's autoregressive transformer). The idea behind TranceptEVE is that sequence and alignment signals are conditionally independent and can be combined as follows:
    $$ \log \mathbb{P}(x_i | x_{1:i-1}) = (1 - \alpha_P )[ (1 - \beta_P) \mathbb{P}_{\text{Transformer}}(x_i | x_{1:i-1}) + \beta_P \mathbb{P}_{\text{MSA}}(x_i | x_{1:i-1}) ]) $$ 
    $$+ \alpha_P \mathbb{P}_{\text{EVE}}(x_i | x_{1:i-1}) $$
    
    \item \textbf{SSemb \citep{blaabjerg_joint_2023}:} SSEmb proposes using a graph neural network to consume protein structure graph and use embeddings from the MSA Transformer model in order to make predictions.
    \item \textbf{SaProt$^*$ \citep{su_saprot_2023}:} SaProt uses Foldseek \citep{van_kempen_fast_2024} to generate protein residue tokens from its 3D structure. It then uses a PLM to process the tokens and compute probabilities of the masked tokens within a context.
    \item \textbf{ESM2 650M \citep{lin_evolutionary-scale_2023}}: Is a protein language model trained on protein sequence datasets using masked language modeling. It uses a BERT-style encoder architecture to learn continuous embedding representations of protein sequences.
    \item \textbf{Ensemble 1 (StructSeq)}: This ensemble described in \citep{paul_combining_nodate} assumes that the probability of a masked token belonging to each amino acid class based on its structure and alignment as well as that from a PLM are mutually independent. Therefore, the three log probabilities can be added to compute the total probability. This model combines log probabilities from ESM inverse folding and TranceptEVE. Unlike other ensembling strategies \citep{kulikova_two_2023}, there is no additional training after combining the base models.
    \item \textbf{Ensemble 2}: We ensemble scores from Tranception, which is a multi-modal model combining PLM and MSA signals and ESM-IF1, which is a structure-based model. We chose this model due to its similarity to Ensemble 1.
    \item \textbf{Ensemble 3}: We ensemble scores from GEMME, which is a MSA-based model, and ESM inverse folding, which is a structure-based model.
    \item \textbf{Ensemble 4}: We ensemble scores from VESPA, which is a PLM, and ESM inverse folding. 
\end{itemize}

\subsection*{Evaluation metrics}
Spearman $\rho$ can be defined as:
    $$\rho = 1 - \frac{6\sum_{i=1}^{N}d_i^2}{N(N^2-1)}$$
    
    where each $d_i$ is a rank difference between $i^{\text{th}}$ prediction and $i^{\text{th}}$ ground truth data point.

    This metric is especially useful when judging how well a model's predictions correlate with a certain quantitative function in a protein. The average value indicates how well the model predicts all protein functions on average.
    
Top 10 Recall can be defined as:
    $$\text{Top-10 Recall} = \frac{\text{ No. of data points that lie in the top 10\% of both predicted and measured values}}{\text{No. of data points in the top 10\% of measured values  }}$$

    The utility of this metric is scenarios where we use a machine learning model to identify mutations in a protein that increase a certain function the most, i.e. identifying the most beneficial mutations. That setting is relevant for protein engineering.

\subsection*{Selection process for disordered proteins} \label{app:idr_selection}

We follow a multi-step process to select proteins containing IDRs from the set of proteins in ProteinGym. First, we select disordered annotations according to protein UniProt IDs present in ProteinGym. Then, we perform a sequence matching operation. We need to do this due to DisProt using protein sequences from UniProt \citep{aspromonte_disprot_2024}. This convention is different from ProteinGym, where each DMS assay is mapped to the sequence used to generate the DMS data \citep{notin_proteingym_2023}. We found that target sequences in ProteinGym can often be different from the UniProt sequences at arbitrary positions.
For example, the sequences for \textalpha-synuclein in UniProt and ProteinGym are of the same length but differ in residues at arbitrary locations throughout the sequence. We detail the sequence alignment procedure below. Lastly, we select mutations within substitution DMS assays in ProteinGym.
\begin{itemize}
    \item \textbf{1$^{\textbf{st}}$ Stage:} We use the DisProt database to query each UniProt ID in ProteinGym and retrieve the set of proteins that have disordered annotations in DisProt.
    \item \textbf{2$^{\textbf{nd}}$ Stage:} 
    \begin{itemize}
        \item We first align the ProteinGym target sequence for each assay to its UniProt sequence. We do this by handling two cases: sequences differing in arbitrary mutations and target sequences that are subsets of UniProt sequences. If target sequences are of the same length as their corresponding UniProt sequences, we treat them as referring to the same sequence. If the target sequence and UniProt sequence differ in length (the case where target sequences are a subset of the UniProt sequence), we find the location of the target sequence within the UniProt sequence.
        \item Using these aligned annotations, we select disordered annotations that lie within the target sequence provided by ProteinGym for each UniProt ID \citep{notin_proteingym_2023}. At the end of this stage, we are left with DMS assays from ProteinGym that contain at least one disordered region.
    \end{itemize}
    \item \textbf{3$^{\textbf{rd}}$ Stage:} We select DMS assays that have mutations in both ordered and disordered regions.
\end{itemize}

Stage 1 yields 58 of the 186 (31\%) unique UniProt IDs in ProteinGym that are proteins that are annotated as having disordered regions in DisProt (\tablename{  \ref{tab:disorder_content}}). These proteins are associated with 69 of the 217 (32\%) ProteinGym DMS substitution assays.

After Stage 2, we obtain 52 of the 186 (28\%) unique UniProt IDs in ProteinGym associated with 63 of the 217 (29\%) of the DMS substitution assays. These assays contain a valid disordered region in \tablename{ \ref{tab:disorder_content}}.
We use the subset of assays obtained after Stage 2 in our analysis in \figurename{ \ref{fig:prot_level_order_vs_dirsorder_spearman}} for benchmarking how different models score proteins which contain disordered regions.

Lastly, in order to benchmark how well models predict functional effects of mutations in ordered versus disordered regions, we select assays that contain mutations in both ordered and disordered regions during Stage 3. After this stage, we are left with 36 unique UniProt IDs corresponding to 43 different assays. The set of assays obtained after Stage 3 are marked with a $^\dagger$ in \tablename{ \ref{tab:disorder_content}} and are used in \figurename{ \ref{fig:order_vs_dirsorder_spearman:contd}}.

\subsection*{Selection process for experimental structures} \label{app:exp_selection}
 We queried the RCSB PDB \citep{bittrich_rcsb_2024} to get structures that exactly match the entire target sequence in ProteinGym. We used the logic to query PDB by performing a search using sequence similarities from the code provided by \citet{boca_predicting_nodate}. 
 
 Our filtering process follows two steps:
 \begin{itemize}
     \item  As our first filtering criteria, we discard proteins that do not have any matching experimental structures. We used an E-value cutoff of 1.0 and identity cutoff of 90\%. We found 82 such experimental structures.  
     \item Our second filtering criteria is to drop any proteins that match to experimental structures that do not contain coordinates for all residues in the ProteinGym target sequence. This is due to the fact that the ESM-IF1 model would not be able to process such structures.
 \end{itemize}

A limitation of our approach is that we did not manually confirm that the selected structures match the functional assays in the ProteinGym benchmark. This can be detrimental to making fitness predictions. For example, if an experimental structure represents a conformation that does not expose an active site, it will be less informative in predicting protein function.

\clearpage

\begin{figure}[htbp]
    \begin{center}
        \includegraphics[width=15cm,height=30cm, keepaspectratio]{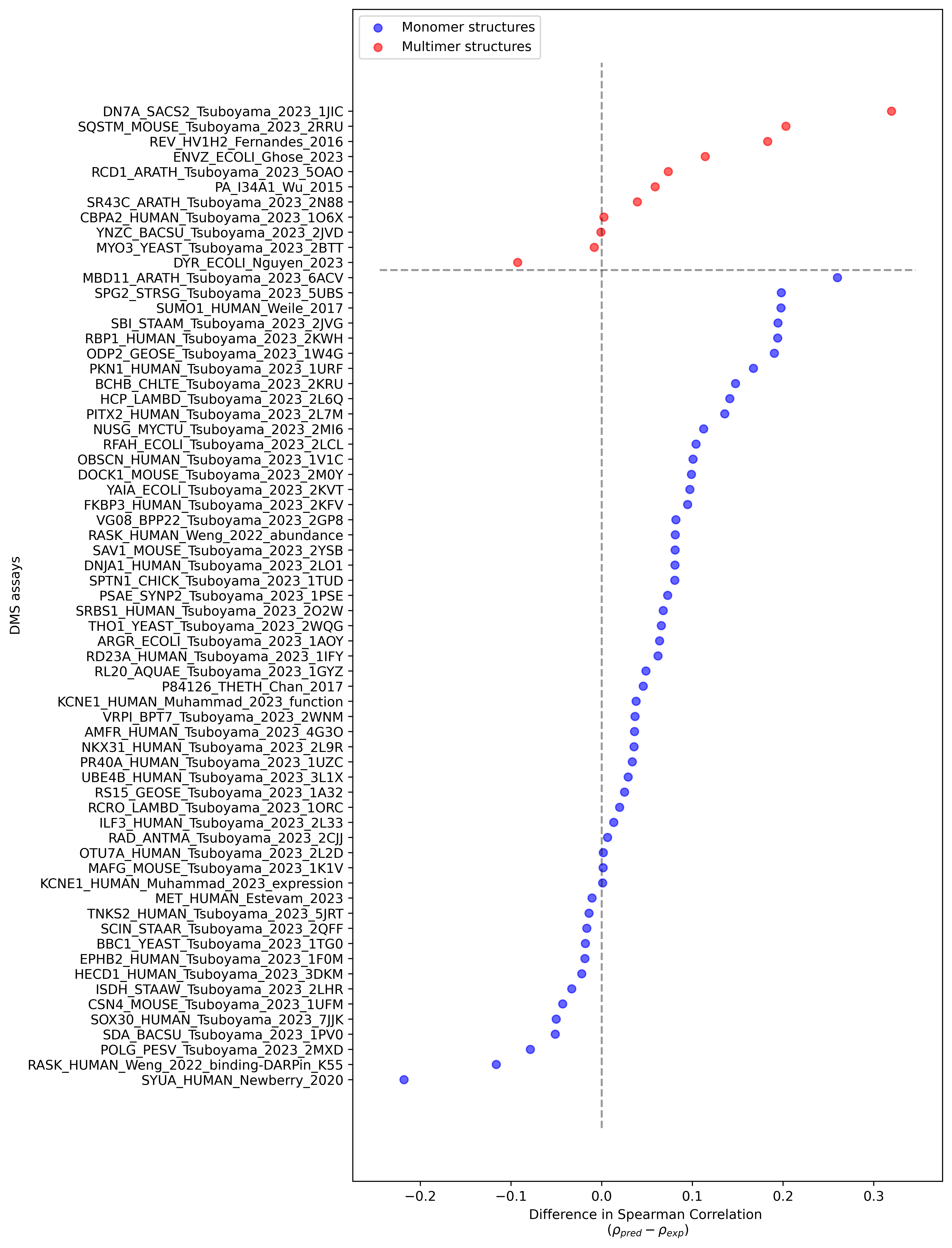}
        \caption{Difference in Spearman correlation between predictions made using predicted and experimental structures. Each datapoint represents a DMS assay for which an experimental structure exists.}
        \label{fig:spearman_diff}
    \end{center}
\end{figure}

\begin{figure}
    \centering
    \includegraphics[width=0.8\linewidth]{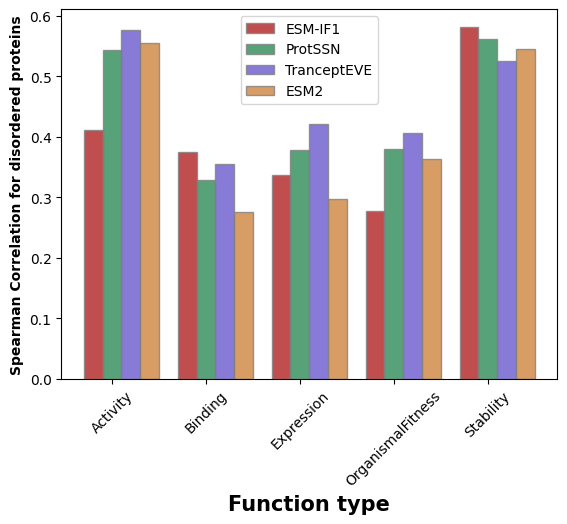}
    \caption{Protein-level Spearman $\rho$ predicting function of proteins comprising some form of disordered regions in the ProteinGym assay target sequence according to the DisProt database.}
    \label{fig:prot_level_order_vs_dirsorder_spearman}
\end{figure}

\begin{figure}[htbp]
    \centering
    \begin{subfigure}{0.45\textwidth}
        \centering
        \includegraphics[width=\linewidth]{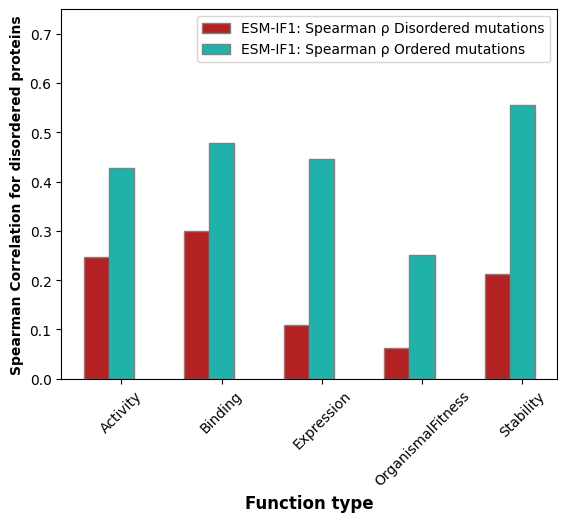}
        \caption{ESM-IF1 spearman $\rho$}
        \label{fig:order_vs_dirsorder_spearman:esm_if}
    \end{subfigure}
    \hfill
    \begin{subfigure}{0.45\textwidth}
        \centering
        \includegraphics[width=\linewidth]{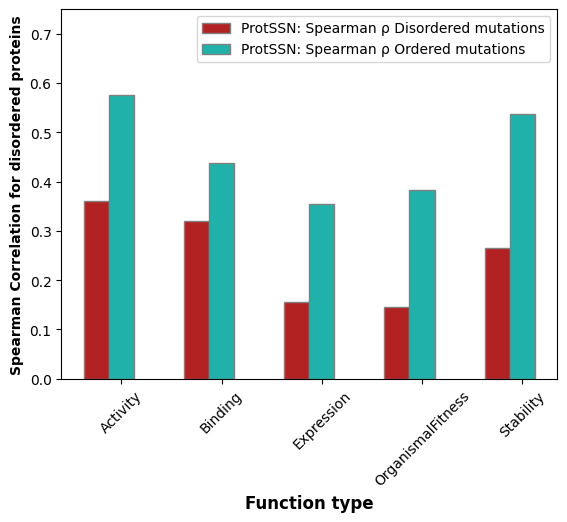}
        \caption{ProtSSN spearman $\rho$}
        \label{fig:order_vs_dirsorder_spearman:protSSN}
    \end{subfigure}

    \vskip\baselineskip 

    \begin{subfigure}{0.45\textwidth}
        \centering
        \includegraphics[width=\linewidth]{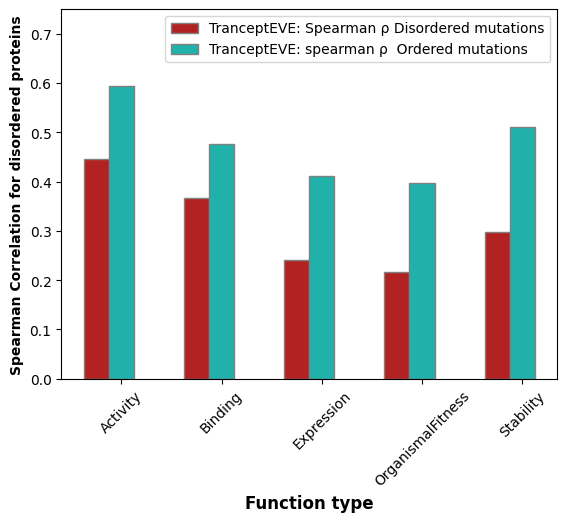}
        \caption{TranceptEVE L spearman $\rho$}
        \label{fig:order_vs_dirsorder_spearman:tranceptEVE}
    \end{subfigure}
    \hfill
    \begin{subfigure}{0.45\textwidth}
        \centering
        \includegraphics[width=\linewidth]{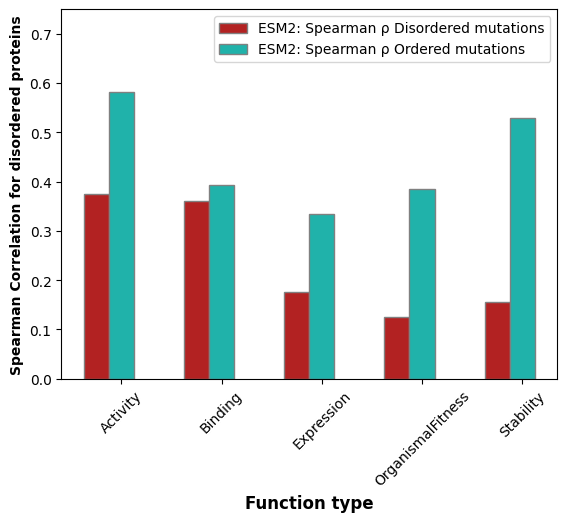}
        \caption{ESM2 650M spearman $\rho$}
        \label{fig:order_vs_dirsorder_spearman:esm2}
    \end{subfigure}

    \vskip\baselineskip 

    \begin{subfigure}{0.45\textwidth}
        \centering
        \includegraphics[width=\linewidth]{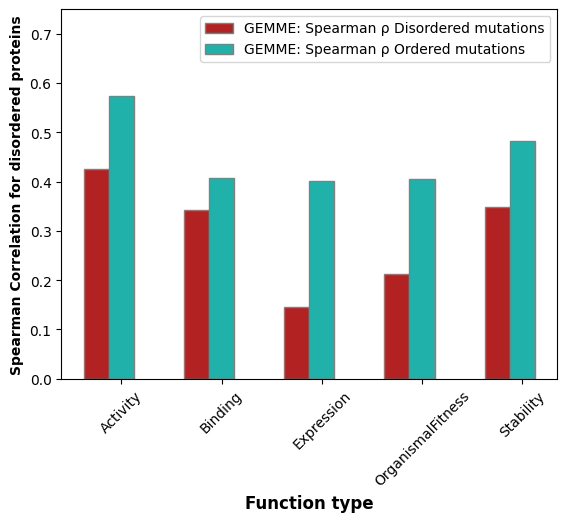}
        \caption{GEMME spearman $\rho$}
        \label{fig:order_vs_dirsorder_spearman:gemme}
    \end{subfigure}
    \hfill
    \begin{subfigure}{0.45\textwidth}
        \centering
        \includegraphics[width=\linewidth]{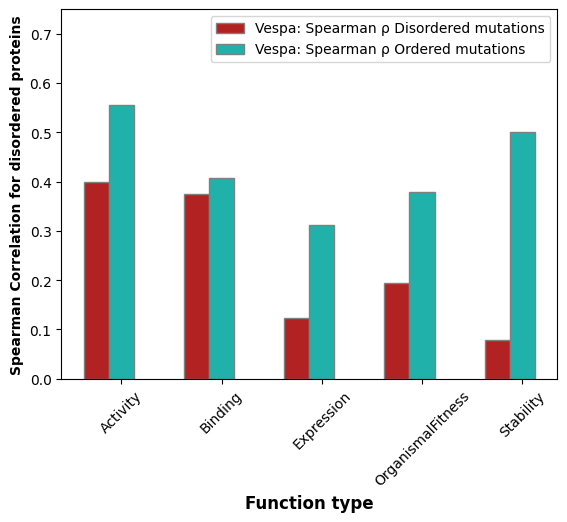}
        \caption{Vespa spearman $\rho$}
        \label{fig:order_vs_dirsorder_spearman:vespa}
    \end{subfigure}
    
    \label{fig:order_vs_dirsorder_spearman}

\end{figure}

\begin{figure}[htbp]
    \centering
    \addtocounter{figure}{-1} 
    \begin{subfigure}{0.45\textwidth}
        \centering
        \includegraphics[width=\linewidth]{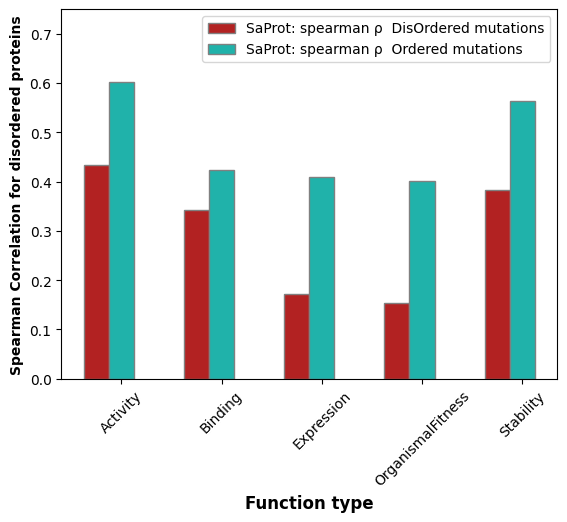}
        \addtocounter{subfigure}{+6} 
        \caption{SaProt 650M spearman $\rho$}
        \label{fig:order_vs_dirsorder_spearman:saProt}
    \end{subfigure}
    \hfill
    \begin{subfigure}{0.45\textwidth}
        \centering
        \includegraphics[width=\linewidth]{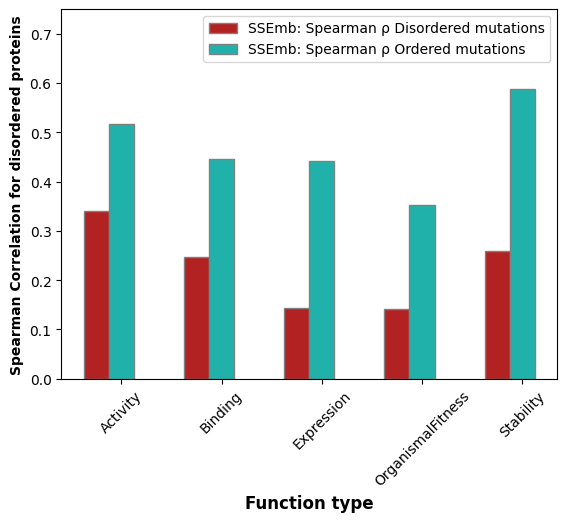}
        \caption{SSEmb spearman $\rho$}
        \label{fig:order_vs_dirsorder_spearman:ssemb}
    \end{subfigure}

    \vskip
    \baselineskip 

    \begin{subfigure}{0.45\textwidth}
        \ContinuedFloat
        \centering
        \includegraphics[width=\linewidth]{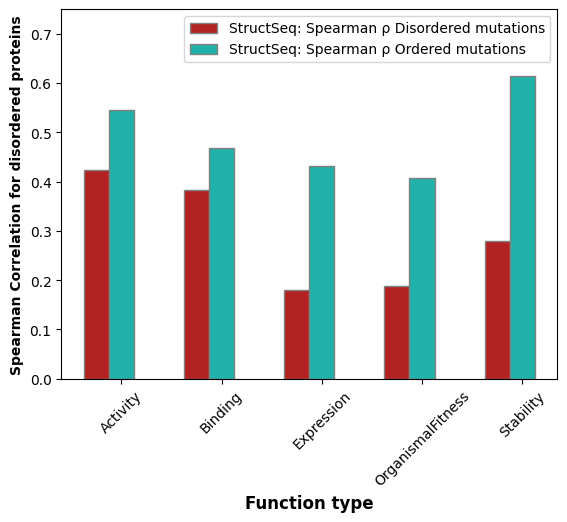}
        \caption{StructSeq Ensemble spearman $\rho$}
        \label{fig:order_vs_dirsorder_spearman:struct2seq}
    \end{subfigure}
    \hfill
    \begin{subfigure}{0.45\textwidth}
        \ContinuedFloat
        \centering
        \includegraphics[width=\linewidth]{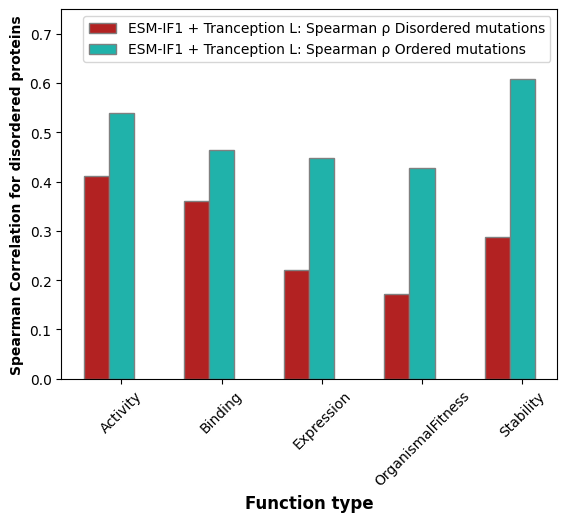}
        \caption{ESM-IF1 + Tranception Ensemble spearman $\rho$}
        \label{fig:order_vs_dirsorder_spearman:ensemble2}
    \end{subfigure}

    \vskip\baselineskip 

    \begin{subfigure}{0.45\textwidth}
        \centering
        \includegraphics[width=\linewidth]{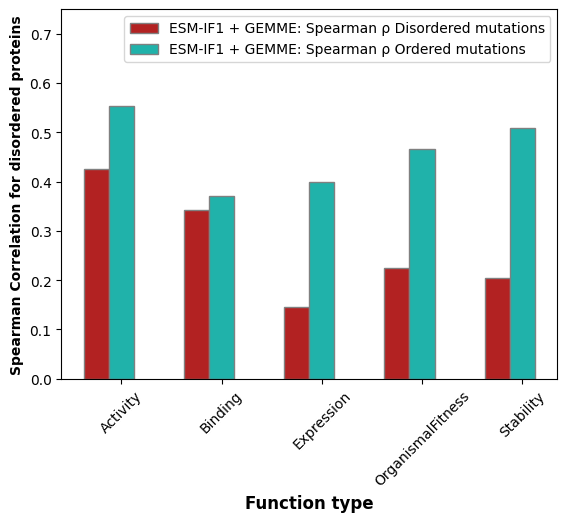}
        \caption{ESM-IF1 + GEMME Ensemble spearman $\rho$}
        \label{fig:order_vs_dirsorder_spearman:ensemble3}
    \end{subfigure}
    \hfill
    \begin{subfigure}{0.45\textwidth}
        \centering
        \includegraphics[width=\linewidth]{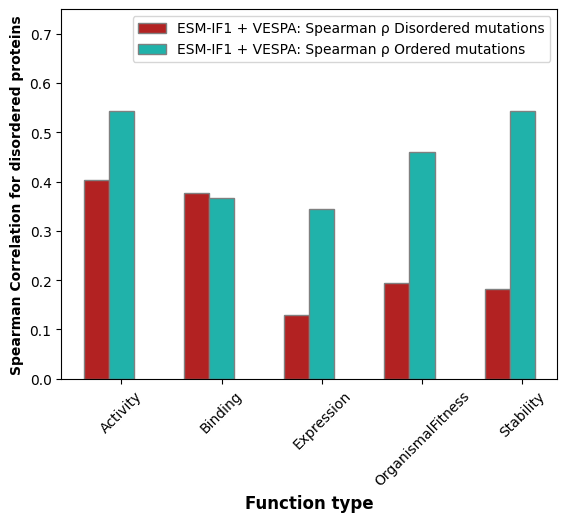}
        \caption{ESM-IF1 + Vespa Ensemble spearman $\rho$}
        \label{fig:order_vs_dirsorder_spearman:ensemble4}
    \end{subfigure}
    \captionsetup{justification=centering} 
    \caption{Spearman correlation averaged across UniProtID and then function type for disordered and ordered regions. These scores were computed across proteins that contain mutations in both disordered and ordered regions.}
    \label{fig:order_vs_dirsorder_spearman:contd}
\end{figure}

\begin{figure}
    \centering
    \begin{subfigure}{0.45\textwidth}
        \centering
        \includegraphics[width=\linewidth]{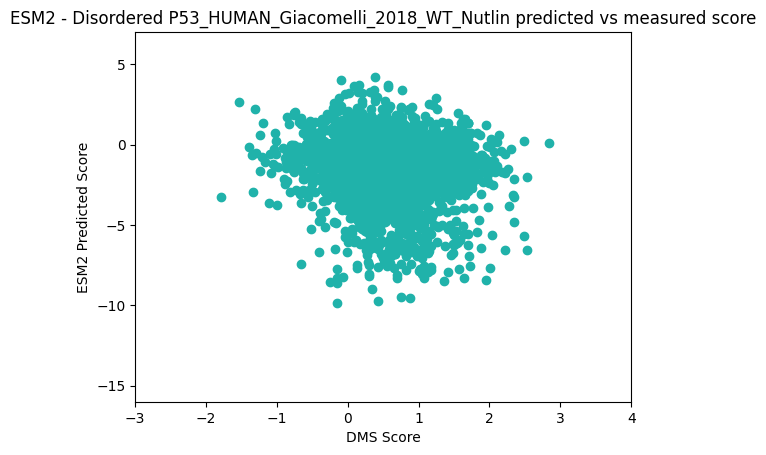}
        \caption{ESM2 predictions for mutations in the disordered regions}
        \label{fig:esm2_order_vs_dirsorder_p53:disorder}
    \end{subfigure}
    \hfill
    \begin{subfigure}{0.45\textwidth}
        \centering
        \includegraphics[width=\linewidth]{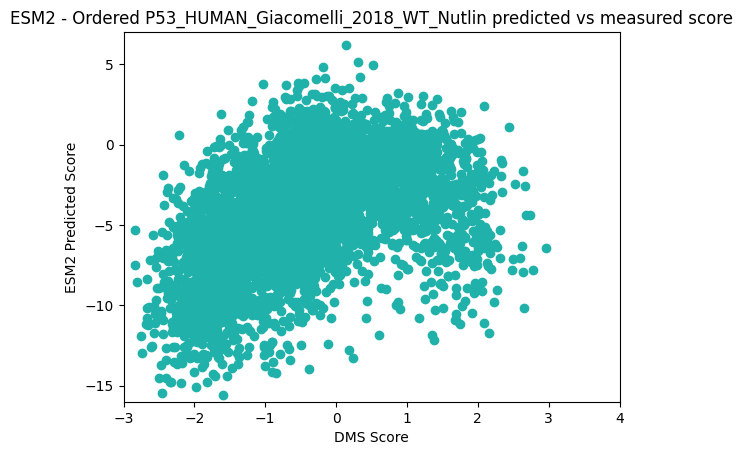}
        \caption{ESM2 predictions for mutations in the ordered regions}
        \label{fig:esm2_order_vs_dirsorder_p53:order}
    \end{subfigure}
    \caption{Predictions made by ESM2 650M on the P53\_HUMAN\_Giacomelli\_2018\_WT\_Nutlin dataset separated by disordered and ordered regions.}
    \label{fig:esm2_order_vs_dirsorder_p53}
\end{figure}

\begin{figure}[htbp]
    \begin{center}
    \includegraphics[width=\linewidth]{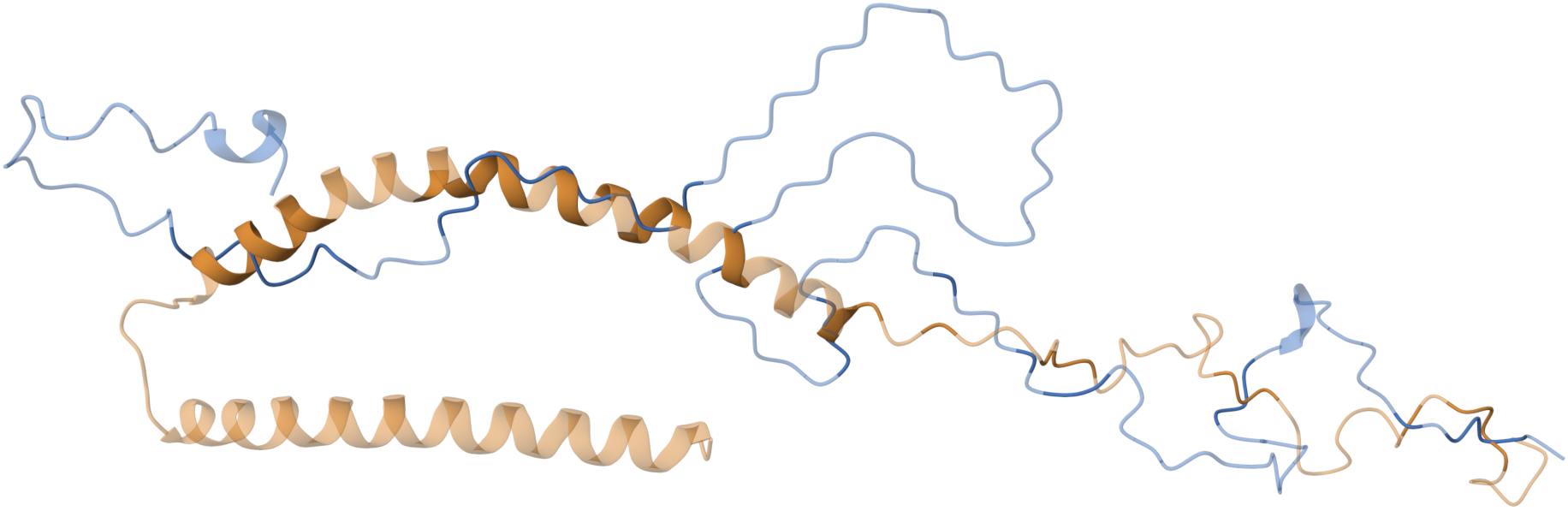}
        \caption{Protein structure alignment between the experimental (orange, PDB 1XQ8 \citep{ulmer_structure_2005}) and predicted (blue) structures for P37840 (human \textalpha-synuclein). The ProteinGym AlphaFold 2 predicted structure resembles an \textalpha-synuclein fibril, PDB 2N0A \citep{tuttle_solid-state_2016}, which is a different conformation than the predicted  \textalpha-synuclein structure in the AlphaFold Protein Structure Database (AF-P37840-F1-v4) \citep{varadi_alphafold_2024}.}
        \label{fig:1XQ8_alignment}
    \end{center}
\end{figure}

\begin{figure}[htbp]
    \begin{center}
    \includegraphics[scale=0.3]{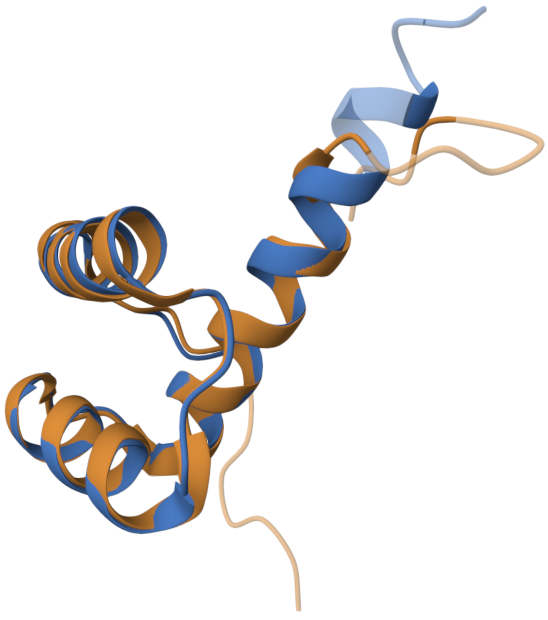}
        \caption{Protein structure alignment between the experimental (orange, PDB 2L9R) and predicted (blue) structures for Q99801 (human NKX3-1). The experimental structure has sequence length of 69 versus 61 for the predicted structure.}
        \label{fig:2L9R_alignment}
    \end{center}
\end{figure}

\begin{figure}[htb]
\begin{adjustwidth}{-2cm}{}
    \begin{center}
        \includegraphics[width=18cm,height=12cm, keepaspectratio]{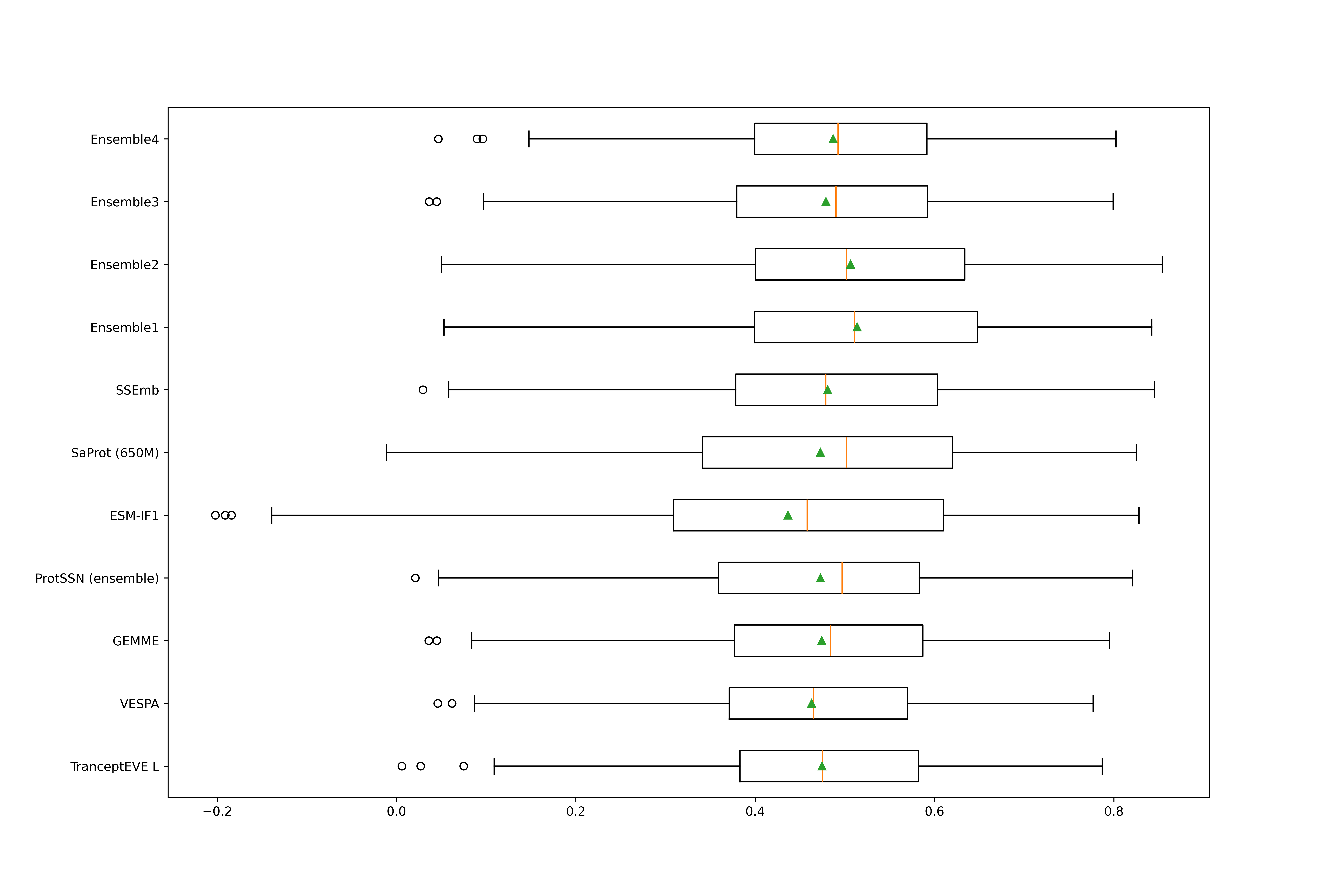}
        \caption{Distribution of Spearman correlation of various models being considered across 216 DMS assays from ProteinGym. Triangles denote the mean correlation. Ensemble1 is the StructSeq method proposed by \citep{paul_combining_nodate}.}
        \label{fig:spearman_distribution}
    \end{center}
    \end{adjustwidth}
\end{figure}

\clearpage

\begin{table}[htbp]
    \begin{center}
    \begin{tabular}{lcccccc}
            \toprule
            \multirow{1}{*}{\textbf{Assay function type}} &
            {Activity} & {Binding} & {Expression} & {Organismal} &
            {Stability} & \textbf{Total}\\
            \midrule \\
            \textbf{Assays with} &
            {43} & {13} & {18} & {76} & {66} & {216}\\
            {\textbf{predicted structures}} & {} & {} & {} & {} & {} & {}\\
            {} & {} & {} & {} & {} & {} & {}\\
            \textbf{Assays with} &
            {3} & {1} & {2} & {6} & {53} & {65}\\
            \textbf{{experimental structures}} & {} & {} & {} & {} & {} & {}\\
            \bottomrule
        \end{tabular}
    \end{center}
    \caption{Count of DMS assays by function type.}
    \label{tab:count_table}
\end{table}

\setlength{\LTleft}{-2.5cm}  
\setlength{\LTright}{\fill} 
\begin{longtable}{llll}
\toprule
\textbf{ProteinGym DMS ID} & \textbf{UniProt} & \textbf{Disorder content (\%)}& \textbf{Disordered} \\
 &  & & \textbf{regions} \\
\midrule
A4\_HUMAN\_Seuma\_2022$^\textbf{\textdagger}$ & P05067 & 5.19 & 671-710\\
ACE2\_HUMAN\_Chan\_2020 & Q9BYF1 & 4.60 & 768-804\\
ADRB2\_HUMAN\_Jones\_2020$^\textbf{\textdagger}$ & P07550 & 17.43 & 341-412\\
B2L11\_HUMAN\_Dutta\_2010\_binding-Mcl-1 & O43521 & 55.56 & 0-40,96-164\\
BRCA1\_HUMAN\_Findlay\_2018$^\textbf{\textdagger}$ & P38398 & 83.20 & 99-1648\\
BRCA2\_HUMAN\_Erwood\_2022\_HEK293T & P51587 & 6.93 & 47-283\\
CALM1\_HUMAN\_Weile\_2017$^\textbf{\textdagger}$ & P0DP23 & 6.71 & 75-84\\
CASP3\_HUMAN\_Roychowdhury\_2020$^\textbf{\textdagger}$ & P42574 & 16.28 & 28-37,172-182,136-156\\
CASP7\_HUMAN\_Roychowdhury\_2020$^\textbf{\textdagger}$ & P55210 & 5.34 & 173-187\\
CATR\_CHLRE\_Tsuboyama\_2023\_2AMI$^\textbf{\textdagger}$ & P05434 & 4.17 & 0-2\\
CBS\_HUMAN\_Sun\_2020$^\textbf{\textdagger}$ & P35520 & 7.26 & 0-39\\
CD19\_HUMAN\_Klesmith\_2019\_FMC\_singles$^\textbf{\textdagger}$ & P15391 & 2.70 & 137-151\\
CUE1\_YEAST\_Tsuboyama\_2023\_2MYX$^\textbf{\textdagger}$ & P38428 & 9.62 & 0-4\\
DLG4\_HUMAN\_Faure\_2021 & P78352 & 9.81 & 0-70\\
DYR\_ECOLI\_Nguyen\_2023$^\textbf{\textdagger}$ & P0ABQ4 & 27.04 & 8-23,62-71,115-131\\
DYR\_ECOLI\_Thompson\_2019$^\textbf{\textdagger}$ & P0ABQ4 & 27.04 & 8-23,62-71,115-131\\
ERBB2\_HUMAN\_Elazar\_2016 & P04626 & 21.35 & 987-1254\\
GAL4\_YEAST\_Kitzman\_2015 & P04386 & 4.31 & 106-143\\
GCN4\_YEAST\_Staller\_2018$^\textbf{\textdagger}$ & P03069 & 47.69 & 0-133\\
HMDH\_HUMAN\_Jiang\_2019$^\textbf{\textdagger}$ & P04035 & 3.15 & 860-887\\
KCNE1\_HUMAN\_Muhammad\_2023\_expression$^\textbf{\textdagger}$ & P15382 & 50.39 & 23-44,71-90,106-128\\
KCNE1\_HUMAN\_Muhammad\_2023\_function$^\textbf{\textdagger}$ & P15382 & 50.39 & 23-44,71-90,106-128\\
MET\_HUMAN\_Estevam\_2023$^\textbf{\textdagger}$ & P08581 & 6.97 & 166-185\\
MK01\_HUMAN\_Brenan\_2016$^\textbf{\textdagger}$ & P28482 & 8.33 & 0-14,174-188\\
MTHR\_HUMAN\_Weile\_2021$^\textbf{\textdagger}$ & P42898 & 1.68 & 160-170\\
NKX31\_HUMAN\_Tsuboyama\_2023\_2L9R & Q99801 & 0 & {Disordered region not}\\
& & & {in target seq.}\\
NPC1\_HUMAN\_Erwood\_2022\_HEK293T$^\textbf{\textdagger}$ & O15118 & 12.83 & 287-332,373-382,604-619,\\
{} & {} & {} & {772-812,959-982,1251-1277}\\
NPC1\_HUMAN\_Erwood\_2022\_RPE1 & O15118 & 12.83 & 287-332,373-382,604-619,\\
{} & {} & {} & {772-812,959-982,1251-1277}\\
NUSA\_ECOLI\_Tsuboyama\_2023\_1WCL & P0AFF6 & 100.0 & 0-68\\
P53\_HUMAN\_Giacomelli\_2018\_Null\_Etoposide$^\textbf{\textdagger}$ & P04637 & 37.66 & 0-92,290-311,360-392\\
P53\_HUMAN\_Giacomelli\_2018\_Null\_Nutlin$^\textbf{\textdagger}$ & P04637 & 37.66 & 0-92,290-311,360-392\\
P53\_HUMAN\_Giacomelli\_2018\_WT\_Nutlin$^\textbf{\textdagger}$ & P04637 & 37.66 & 0-92,290-311,360-392\\
P53\_HUMAN\_Kotler\_2018$^\textbf{\textdagger}$ & P04637 & 37.66 & 0-92,290-311,360-392\\
PABP\_YEAST\_Melamed\_2013 & P04147 & 14.56 & 418-501\\
PAI1\_HUMAN\_Huttinger\_2021$^\textbf{\textdagger}$ & P05121 & 3.98 & 354-369\\
PA\_I34A1\_Wu\_2015$^\textbf{\textdagger}$ & P03433 & 3.63 & {371-396}\\
PIN1\_HUMAN\_Tsuboyama\_2023\_1I6C$^\textbf{\textdagger}$ & Q13526 & 15.38 & 33-38\\
POLG\_DEN26\_Suphatrakul\_2023 & P29990 &  0 & {Disordered region not}\\
& & & {in target seq.}\\
POLG\_HCVJF\_Qi\_2014 & Q99IB8 & 6.73 & 2223-2316,2330-2439\\
PPARG\_HUMAN\_Majithia\_2016$^\textbf{\textdagger}$ & P37231 & 59.21 & 8-210,237-297,442-476\\
PRKN\_HUMAN\_Clausen\_2023$^\textbf{\textdagger}$ & O60260 & 16.13 & 72-98,108-141,377-390\\
PTEN\_HUMAN\_Matreyek\_2021$^\textbf{\textdagger}$ & P60484 & 18.61 & 285-308,352-402\\
PTEN\_HUMAN\_Mighell\_2018$^\textbf{\textdagger}$ & P60484 & 18.61 & 285-308,352-402\\
R1AB\_SARS2\_Flynn\_2022 & P0DTD1 & 0 & {Disordered region not}\\
& & & {in target seq.}\\
RAF1\_HUMAN\_Zinkus-Boltz\_2019 & P04049 & 3.09 & 235-254\\
RASH\_HUMAN\_Bandaru\_2017 & P01112 & 10.58 & 169-188\\
RASK\_HUMAN\_Weng\_2022\_abundance$^\textbf{\textdagger}$ & P01116 & 17.55 & 59-69,166-187\\
RASK\_HUMAN\_Weng\_2022\_binding-DARPin\_K55$^\textbf{\textdagger}$ & P01116 & 17.55 & 59-69,166-187\\
RCD1\_ARATH\_Tsuboyama\_2023\_5OAO$^\textbf{\textdagger}$ & Q8RY59 & 1.75 & 0-0\\
RCRO\_LAMBD\_Tsuboyama\_2023\_1ORC$^\textbf{\textdagger}$ & P03040 & 15.87 & 53-62\\
RD23A\_HUMAN\_Tsuboyama\_2023\_1IFY$^\textbf{\textdagger}$ & P54725 & 9.09 & 40-43\\
RFAH\_ECOLI\_Tsuboyama\_2023\_2LCL$^\textbf{\textdagger}$ & P0AFW0 & 12.73 & 0-6\\
RL20\_AQUAE\_Tsuboyama\_2023\_1GYZ$^\textbf{\textdagger}$ & O67086 & 49.15 & 0-0, 31-58\\
SHOC2\_HUMAN\_Kwon\_2022$^\textbf{\textdagger}$ & Q9UQ13 & 14.60 & 0-84\\
SPIKE\_SARS2\_Starr\_2020\_binding & P0DTC2 & 39.59 & 0-25,66-79,141-163,172-184,\\
{} & {} & {} & {245-261,318-540,620-639,}\\
{} & {} & {} & {672-686,827-852,1146-1272}\\
SPIKE\_SARS2\_Starr\_2020\_expression & P0DTC2 & 39.59 & 0-25,66-79,141-163,172-184,\\
{} & {} & {} & {245-261,318-540,620-639,}\\
{} & {} & {} & {672-686,827-852,1146-1272}\\
SRBS1\_HUMAN\_Tsuboyama\_2023\_2O2W & Q9BX66 & 0 & {Disordered region not}\\
& & & {in target seq.}\\
SRC\_HUMAN\_Ahler\_2019 & P12931 & 15.67 & 0-83\\
SRC\_HUMAN\_Chakraborty\_2023\_binding-DAS\_25uM & P12931 & 15.67 & 0-83\\
SRC\_HUMAN\_Nguyen\_2022 & P12931 & 15.67 & 0-83\\
SUMO1\_HUMAN\_Weile\_2017$^\textbf{\textdagger}$ & P63165 & 17.82 & 0-17\\
SYUA\_HUMAN\_Newberry\_2020 & P37840 & 100.0 & 0-139\\
TADBP\_HUMAN\_Bolognesi\_2019 & Q13148 & 36.71 & 262-413\\
TAT\_HV1BR\_Fernandes\_2016 & P04610 & 100.0 & 0-85\\
UBE4B\_HUMAN\_Tsuboyama\_2023\_3L1X$^\textbf{\textdagger}$ & O95155 & 1.45 & 0-0\\
UBR5\_HUMAN\_Tsuboyama\_2023\_1I2T & O95071 & 0 & Disordered region not \\
{} & {} & {} & {in target seq.}\\
VILI\_CHICK\_Tsuboyama\_2023\_1YU5$^\textbf{\textdagger}$ & P02640 &  3.08 & 0-1\\
YAP1\_HUMAN\_Araya\_2012$^\textbf{\textdagger}$ & P46937 & 24.21 & 49-170\\
YNZC\_BACSU\_Tsuboyama\_2023\_2JVD & O31818 & 0 & Disordered region not \\
{} & {} & {} & {in target seq.}\\
\bottomrule
 \caption{DisProt disorder content for ProteinGym proteins and the associated ProteinGym substitution DMS assays.
 The disorder content and 0-indexed disordered regions pertain only to the subset of the protein sequence covered by the DMS assay (target sequence). Disordered regions for proteins with disordered annotation outside of the target sequence are marked as such. Additionally, DMS assays used in \figurename{ \ref{fig:order_vs_dirsorder_spearman:contd}} containing mutations in both ordered and disordered regions marked with ($^\dagger$).} 
\label{tab:disorder_content}
\end{longtable}

\begin{table}[htbp]
    \begin{center}
        \begin{tabular}{lcccccc}
        \toprule
         \multirow{1}{*}{\textbf{Assay function type}} &
            {Activity} & {Binding} & {Expression} & {Organismal} &
            {Stability} & \textbf{Total}\\
        \midrule \\
        \textbf{DMS ID Count} &
            {9} & {4} & {4} & {15} & {11} & \textbf{43}\\
        \bottomrule
        \end{tabular}
    \end{center}
    \caption{Count of DMS IDs corresponding to disordered proteins identified by 36 unique UniProt IDs that have mutations in both ordered and disordered regions.}
    \label{tab:count_disordered_proteins}
\end{table}

\begin{table}[htbp]
\begin{adjustwidth}{-2cm}{}
        \begin{tabular}{@{\hspace{-0.3cm}} l c cccccc @{\hspace{-0.7cm}} c} 
            \toprule
            \multirow{2}{*}{\textbf{Model}} &
            \multirow{2}{*}{\textbf{Modality}} &
            \multicolumn{5}{c}{\textbf{Top 10 Recall (R) by Assay Function}} &
            \multirow{2}{*}{\textbf{Avg. Top 10 R}} \\
            & &  &  &  &  & \\
            & & {Activity} & {Binding} & {Expression} & {Organismal} & {Stability} \\
            & & {} & {} & {} & {Fitness} & {} \\
            \midrule
            ESM-IF1$^{*}$ & Structure & 0.180 & 0.207 & 0.205 & 0.173 & \textcolor{red!90}{0.344} & 0.222 & \\
            GEMME$^{*}$ & MSA & 0.194 & 0.200 & 0.198 & 0.215 & 0.233 & 0.208 & \\
            VESPA$^{*}$ & PLM & 0.175 & 0.192 & 0.184 & 0.208 & 0.242 & 0.200 & \\
            ProtSSN$^{*}$ (ensemble) & Structure + PLM & 0.187 & 0.197 & 0.220 & 0.191 & \textcolor{olive!90}{0.337} & 0.227 & \\
            SSemb & Structure + MSA & \textcolor{blue!90}{0.198} & 0.222 & 0.204 & 0.215 & 0.334 & \textcolor{blue!90}{0.235} & \\
            TranceptEVE L$^{*}$ & MSA + PLM & 0.197 & 0.221 & \textcolor{blue!90}{0.224} & \textcolor{red!90}{0.229} & 0.278 & 0.230 & \\
            SaProt (650M)$^{*}$ & Structure + PLM & 0.187 & \textcolor{olive!90}{0.232} & \textcolor{red!90}{0.236} & 0.175 & 0.331 & 0.232 & \\
             \midrule
            {Ensemble 1}  & Structure + & & &  &  & & & \\
            StructSeq & MSA + PLM & \textcolor{red!90}{0.208} & \textcolor{red!90}{0.238} & \textcolor{olive!90}{0.232} & \textcolor{blue!90}{0.221} & \textcolor{blue!90}{0.335} & \textcolor{red!90}{0.247} & \\
            \midrule
            Ensemble 2  & Structure + & & &  &  & & & \\
            ESM-IF1, Tranception & MSA + PLM & \textcolor{olive!90}{0.206} & \textcolor{blue!90}{0.224} & 0.221 & \textcolor{blue!90}{0.221} & 0.328 & \textcolor{olive!90}{0.240} & \\
            \midrule
            Ensembled 3 & Structure + & & &  &  & & & \\
            ESM-IF1, GEMME & MSA & {0.197} & {0.199} & 0.198 & \textcolor{olive!90}{0.223} & 0.244 & {0.212} & \\
            \midrule
            Ensemble 4 & Structure + & & &  &  & & & \\
            ESM-IF1, VESPA & PLM & {0.184} & {0.194} & 0.186 & {0.213} & 0.278 & {0.211} & \\
            \bottomrule
        \end{tabular}
    \caption{Top 10 Recall of predictions made by different models sorted according to function type of each DMS assay. Scores for model names annotated with an asterisk ($^{*}$) were calculated using their predictions provided in the ProteinGym benchmark \citep{notin_proteingym_2023}. Scores highlighted in red, olive, blue represent the \textcolor{red!90}{1$^\text{st}$}, \textcolor{olive!90}{2$^\text{nd}$}, and \textcolor{blue!90}{3$^\text{rd}$} highest score for each group. Modalities represented in the table are protein language models (PLMs), structure-based models (Structure), Multiple Sequence Alignment (MSA).}
    \label{tab:recall_table}
    \end{adjustwidth}
\end{table}

\end{document}